\documentclass[useAMS,usenatbib,psfig]{mn2e}
\usepackage{psfig}
\usepackage{times}
\usepackage{color}
\usepackage{ulem}

\title[A toy model for X-ray variability of AGN] {A toy model for X-ray spectral variability of active galactic nuclei}
\author[X. Cao \& J.-X. Wang]
{Xinwu Cao$^{1,2}$ and Jun-Xian Wang$^3$\\
$^1$ Key Laboratory for
Research in Galaxies and Cosmology, Shanghai Astronomical
Observatory, Chinese Academy of Sciences, \\80
Nandan Road, Shanghai, 200030, China; E-mail: cxw@shao.ac.cn\\
$^2$ Department of Astronomy and Institute of Theoretical Physics and Astrophysics, Xiamen University, Xiamen, Fujian, 361005, China\\
$^3$ Key Laboratory for Research in Galaxies and Cosmology,
Department of Astronomy, University of Science and Technology of
China, \\Hefei, Anhui 230026, China  }

\date{Accepted 2014 June 24. Received 2014 June 17; in original form 2014 May 14}

\pagerange{\pageref{firstpage}--\pageref{lastpage}} \pubyear{}

\begin{document}

\maketitle \label{firstpage}

\begin{abstract}
The long term X-ray spectral variability of ten active galactic
nuclei (AGN) shows a positive spectral index-flux correlation for
each object \citep*[][]{2009MNRAS.399.1597S}. {An inner ADAF may
connect to a thin disc/corona at a certain transition radius, which
are responsible for hard X-ray emission in AGN.} The ADAF is hot and
its X-ray spectrum is hard, while the corona above the disc is
relatively cold and its X-ray spectrum is therefore soft. The
radiation efficiency of the ADAF is usually much lower than that of
the thin disc. The increase of the transition radius may lead to
decreases of the spectral index (i.e., a hard spectrum) and the
X-ray luminosity even if the accretion rate is fixed, and vice
versa. We propose that such X-ray variability is caused by the
change of the transition radius. Our model calculations can
reproduce the observed index-flux correlations, if the transition
radius fluctuates around an equilibrium position, and the radiation
efficiency of ADAFs is $\sim 5$ per cent of that for a thin disc.
The average spectral index-Eddington ratio correlation in the AGN
sample can also be reproduced by our model calculations, if the
equilibrium transition radius increases with decreasing mass
accretion rate.
\end{abstract}


\section{Introduction}\label{intro}

The X-ray emission of luminous active galactic nuclei (AGN) is
believed from the inner region of the accretion disc surrounding a
massive black hole \citep*[e.g.,][]{1994ApJ...432L..95H}. In the
accretion disc model, the observed optical/UV continuum spectra are
supposed to originate from the accretion disc
\citep*[][]{1978Natur.272..706S,1982ApJ...254...22M}, while the hard
X-ray emission is from the corona above the disc
\citep*[][]{1979ApJ...229..318G,1991ApJ...380L..51H,1993ApJ...413..507H}.
The gravitational energy of the gas is predominantly released in the
accretion disc, and a fraction of it is transported into the corona,
and therefore the electrons in the corona are heated up to high
temperatures $\sim 10^9$~K
\citep*[][]{2008MNRAS.390..227D,1999MNRAS.304..809D,2001MNRAS.328..958M,2002MNRAS.332..165M,2003ApJ...587..571L,2009MNRAS.394..207C}.
The corona is optically thin, and only a small fraction of the soft
photons in optical/UV bands are inverse Compton scattered to hard
X-ray bands by the hot electrons in the corona. This is the most
favorite scenario for the power law X-ray continuum ubiquitously
observed in luminous AGN
\citep*[][]{1979ApJ...229..318G,1991ApJ...380L..51H,1993ApJ...413..507H}.

The observational features of low luminosity AGN (with Eddington
ratio lower than $\sim 0.01$) are different from their luminous
counterparts, most of which lack of optical/UV bumps in their
continuum spectra and broad emission lines \citep*[see][for a
review]{2008ARA&A..46..475H}. The X-ray spectral indices of low
luminosity AGN are systematically lower than luminous AGN
\citep*[e.g.,][]{1999ApJ...526L...5L}. The statistic study on the
optical/UV/X-ray emission shows evidence that the central engines in
low luminosity AGN are different from those in luminous counterparts
\citep*[][]{2011ApJ...739...64X}. It was suggested that advection
dominated accretion flows (ADAF) are present in low luminosity AGN
\citep*[see][for a review]{2002luml.conf..405N}, which are different
from the standard thin accretion discs in luminous AGN. The radial
velocity of ADAFs is high, typically several tens per cent of the
Keplerian velocity, and their gas density is low, which leads to low
radiation efficiency \citep*[][]{1995ApJ...452..710N}. The typical
electron temperature of the ADAFs can be about one order of
magnitude higher than the temperature of the electrons in the corona
above the disc \citep*[e.g.,][]{2010ApJ...716.1423X}, and the X-ray
spectra of ADAFs are naturally harder than those of the corona in
luminous AGN.

The X-ray emission provides useful clues on the physics of central
engines in AGN. Both the hard X-ray spectral index and the fraction
of X-ray to bolometric luminosity $L_{\rm X}/L_{\rm bol}$ are found
to be correlated with the Eddington ratio for a luminous AGN sample
\citep*[][]{2004ApJ...607L.107W}. These two correlations can be
explained by the accretion disc corona model
\citep*[][]{2009MNRAS.394..207C}.
Unlike luminous AGN, \citet{2009MNRAS.399..349G} found that the
X-ray spectral index is anti-correlated with the Eddington ratio for
a low luminosity AGN sample, which is confirmed by some other works
\citep*[][]{2011A&A...530A.149Y,2012A&A...539A.104Y,2012MNRAS.424.1327E}.
Similar features are also observed in some X-ray binaries
\citep*[e.g.,][]{1999MNRAS.303L..11Z,2004MNRAS.351..791Z,2008ApJ...682..212W,2014arXiv1404.5316C,2014arXiv1404.5317D}.
The optical depth of Compton scattering in ADAF increases with the
dimensionless mass accretion rate $\dot{m}$
($\dot{m}=\dot{M}/\dot{M}_{\rm Edd}$), and therefore the spectral
index decreases with $\dot{m}$. The anti-correlation between
spectral index and Eddington ratio can be qualitatively explained
with the ADAF model \citep*[][]{2009MNRAS.399..349G}.

It is known that many AGN exhibit variability in different time
scales. The long term X-ray spectral variability of ten active
galactic nuclei (AGN) has been studied with the data of
\textit{Rossi X-ray Timing Explorer (RXTE)} observations {over
$\sim$ 10 years}, in which a similar spectral index-flux correlation
is found for {all objects (except for NGC 5548)} over a large range
of flux \citep*[][]{2009MNRAS.399.1597S}. {Such X-ray ``softer when
brighter" behavior can not be interpreted with X-ray absorption by a
single medium whose ionization level follows the continuum flux
variations. The model of a constant reflection component plus a
power law continuum with constant spectral slope and variable flux
is also disfavored, indicating the intrinsic continuum is variable
both in flux and in shape
\citep*[][]{2009MNRAS.399.1597S}.
}

The accretion mode transition is mainly triggered by the
dimensionless mass accretion rate $\dot{m}$, which is sensitive to
the value of the viscosity parameter $\alpha$ \citep*[see][for
reviews]{2002luml.conf..405N,2014arXiv1401.0586Y}. Although the
detailed physics of accretion mode transition is still unclear, it
may be regulated by the processes in the accretion disc corona
systems, i.e., the evaporation of the disc and/or condensation of
the hot gas in the corona
\citep*[][]{1999A&A...348..154M,1999ApJ...527L..17L}. {An
alternative scenario for accretion mode transition being triggered
by the thermal instability was suggested by
\citet{1995ApJ...438L..37A}.} It was indeed found that a small
fraction of low luminosity AGN have optical bumps in their continuum
spectra with hard X-ray spectra, of which the multi-band spectral
energy distributions can be well reproduced by the ADAF$+$standard
thin accretion disc model
\citep*[e.g.,][]{1999ApJ...525L..89Q,2004ApJ...612..724Y,2009RAA.....9..401X}.
This provides evidence for accretion mode transition in AGN.
{\citet{1999MNRAS.303L..11Z} proposed that a hot plasma (probably an
ADAF) surrounded by a cold disc can reproduce the observed
correlation between the X-ray spectral index and the reflection
parameter $R=\Omega/2\pi$ ($\Omega$ is the solid angle subtended by
the reflector) in Seyfert AGNs and the hard state of X-ray binaries.
In this scenario, the power of soft photons from the cold disc
increases with decreasing of the inner radius of the cold disc,
which leads to thermal Comptonization amplification factor $A$
decreasing and softer X-ray spectra.}

In this work, we suggest that the variation of the transition radius
between the ADAF and thin disc/corona is responsible for the
observed X-ray variability in AGNs. In Sect. \ref{model}, we
describe the model calculations. The results and discussion are in
Sects. \ref{results} and \ref{discussion}.

\section{Model}\label{model}


We consider that the black hole is surrounded by an ADAF accreting
at a critical rate $\dot{m}_{\rm crit}$, which connects to a thin
accretion disc with corona at a transition radius $R_{\rm tr}$. The
observed X-ray emission originates from the inner ADAF and the
corona above the disc in the outer region. It was found that the
X-ray spectral index is correlated with the Eddington ratio for
bright AGNs. The photon spectral index $\Gamma\sim 2-2.5$ for the
sources with Eddington ratio $\ga 0.1$, while it decreases to $\sim
1$ if Eddington ratio $\sim 0.01$
\citep*[]{2004ApJ...607L.107W,2006ApJ...646L..29S,2008ApJ...682...81S}.
For low luminosity AGNs, the X-ray spectra are systematically harder
than those of bright AGNs. Their X-ray spectral index is
anti-correlated to the Eddington ratio. The photon spectral index
$\Gamma\sim 1$ for the low luminosity AGNs at high Eddington ratio
end \citep*{2011A&A...530A.149Y}. The observed correlations can be
reproduced by the accretion corona/ADAF models
\citep*{2009MNRAS.394..207C,2013ApJ...764....2Q}. In this work, we
do not intend to get into the complexity of the detailed physics of
these models, and we adopt the X-ray spectral indices of the corona
and ADAF as input model parameters. The location of the transition
radius between the inner ADAF and the thin disc corona is still
unclear, which is believed to be related to values of the disc
parameters, such as, the viscosity parameter $\alpha$, mass
accretion rate $\dot{m}$, and even the magnetic field strength in
the disc
\citep*[e.g.,][]{1999A&A...348..154M,1999ApJ...527L..17L,2001ApJ...546..966K,2012ApJ...759...65T}.
It is known that the radiation efficiency of ADAFs is significantly
lower than that of thin accretion discs, which means that the
radiation power from an ADAF$+$thin disc/corona system varies with
transition radius even if the mass accretion rate remains constant.
We assume that the observed X-ray variability in AGNs is mainly
caused by the transition radius fluctuation around the equilibrium
position.

We use a correction factor $f_{\rm cor}$ to relate the bolometric
luminosity $L_{\rm bol}$ to the X-ray luminosity $L_{\rm X,2-10keV}$
in 2$-$10~keV,
\begin{equation}
L_{\rm bol}=f_{\rm cor}L_{\rm X,2-10keV}. \label{f_cor_1}
\end{equation}
A typical value of $f_{\rm cor}^{\rm b}=30$ is adopted for bright
AGN \citep*[][]{2012MNRAS.422..478R}. The relative contribution of
X-ray photons to the spectral energy distribution (SED) of low
luminosity AGN is higher than their high luminosity counterparts.
\citet{1999ApJ...516..672H} found that $L_{\rm bol}/L_{\rm X}=3-17$
for a small sample of low luminosity AGN with well measured SED. The
SED of low-luminosity AGN can be well fitted with the ADAF model. In
this work, we adopt $f_{\rm cor}^{\rm f}=10$ for the radiation from
ADAFs.

For luminous AGN, the bolometric luminosity of the accretion disc
with corona is
\begin{equation}
L_{\rm bol}=\eta_{\rm rad}\dot{M}c^2, \label{l_bol_1}
\end{equation}
where $\dot{M}$ is the mass accretion rate, and $\eta_{\rm rad}$ is
the radiation efficiency. For radiatively efficient accretion discs,
the radiation efficiency $\eta_{\rm rad}$ is in the range of $\sim
0.05-0.3$, which is a function of the black hole spin parameter $a$.
For simplicity, we limit our calculations in the Newtonian frame.
The radiation efficiency is
\begin{equation}
\eta_{\rm rad}={\frac {1}{2r_{\rm in}}}, \label{eta_rad_1}
\end{equation}
where $r_{\rm in}=R_{\rm in}/R_{\rm g}$ ($R_{\rm g}=GM_{\rm
bh}/c^2$) is the inner radius of a radiatively efficient accretion
disc.

In the case of the accretion disc corona truncated at radius $R_{\rm
tr}$, its X-ray luminosity is
\begin{equation}
L_{\rm X,2-10keV}^{\rm cor}\simeq {\frac {\eta_{\rm
rad}\dot{M}c^2}{f_{\rm cor}^{\rm b}}}\left({\frac {R_{\rm
in}}{R_{\rm tr}}}\right)={\frac {\dot{M}c^2}{2r_{\rm tr}f_{\rm
cor}^{\rm b}}}, \label{l_x_cor1}
\end{equation}
where equation (\ref{eta_rad_1}) is used, and $r_{\rm tr}=R_{\rm
tr}/R_{\rm g}$.

The radiation efficiency of the inner ADAF within $R\le R_{\rm tr}$
is much lower than that of a standard thin accretion disc. The
precise value of the efficiency is still quite uncertain, which is
sensitive to the mass accretion rate $\dot{m}$ and the viscosity
parameter $\alpha$ \citep*[e.g.,][]{2010ApJ...716.1423X}. Assuming
$\eta_{\rm rad}^{\rm ADAF}=f_{\eta}^{\rm ADAF}\eta_{\rm rad}$
($f_{\eta}^{\rm ADAF}<1$), we have
\begin{displaymath}
L_{\rm X,2-10keV}^{\rm ADAF}\simeq {\frac {f_{\eta}^{\rm
ADAF}\eta_{\rm rad}\dot{M}c^2}{f_{\rm cor}^{\rm f}}}\left(1-{\frac
{R_{\rm in}}{R_{\rm tr}}}\right)
\end{displaymath}
\begin{equation}
~~~~~~~~~~~~~~~~~~~~~~~~~={\frac {f_{\eta}^{\rm
ADAF}\dot{M}c^2}{2r_{\rm in}f_{\rm cor}^{\rm f}}}\left(1-{\frac
{r_{\rm in}}{r_{\rm tr}}}\right), \label{l_x_adaf1}
\end{equation}
for an ADAF truncated at $r_{\rm tr}$. For the accretion flows
surrounding non-rotating black holes, $R_{\rm in}=6R_{\rm g}$. The
X-ray luminosity of the ADAF$+$thin disc corona system in 2$-$10~keV
is
\begin{equation}
L_{\rm X,2-10keV}=L_{\rm X,2-10keV}^{\rm ADAF}+L_{\rm
X,2-10keV}^{\rm cor}. \label{l_x_tot1}
\end{equation}
In this work, we adopt the critical mass accretion rate
$\dot{m}_{\rm crit}$ as an input parameter in the model
calculations. The hard X-ray luminosity and spectral shape of
ADAF$+$thin disc corona system changing with the transition radius
$r_{\rm tr}$ can be calculated with equations
(\ref{l_x_cor1})-(\ref{l_x_tot1}), provided the spectral indices of
the X-ray spectra from the ADAF and the outer thin disc corona are
known.

\section{Results}\label{results}


\begin{figure}
\centerline{\psfig{figure=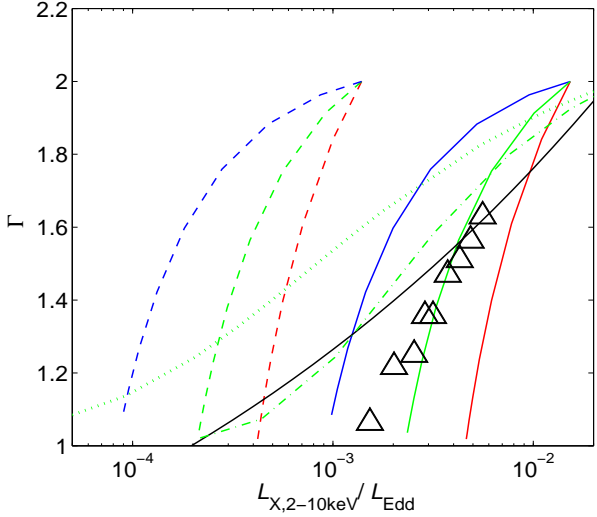,width=8.0cm,height=7.0cm}}
\caption{The relations of the photon spectral index with the
Eddington ratio for different values of parameters. The solid
coloured lines represent the cases with $\dot{m}=0.55$, while the
dashed lines are for $\dot{m}=0.05$. The coloured lines indicate the
results with different values of radiation efficiency for ADAFs,
$f_{\eta}^{\rm ADAF}=0.1$ (red), $0.05$ (green), and $0.02$ (blue).
The triangles are the binned spectral index-flux data of NGC~3516,
{which are shifted to right by $\sim0.5$ dex compared with those in
\citet{2009MNRAS.399.1597S} due to the updated black hole mass
measurement}. The black solid line indicates the best-fitting
correlation between average spectral index and mass accretion rate
{derived from the AGN sample with the updated black hole mass data
(see the text in Sect. \ref{results}). The results with varying mass
accretion rate model calculations are plotted as the dotted line
($p=1$, see equation \ref{r_tr_mdot1}) and the dash-dotted line
($p=2$). } } \label{eddrat_gam}
\end{figure}


\begin{figure}
\centerline{\psfig{figure=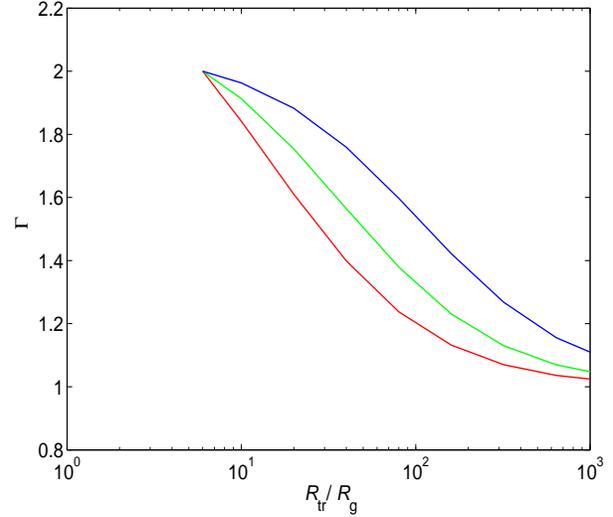,width=8.0cm,height=7.0cm}}
\caption{The photon spectral indices as functions of transition
radius $r_{\rm tr}$.  The coloured lines indicate the results with
different values of radiation efficiency for ADAFs, $f_{\eta}^{\rm
ADAF}=0.1$ (red), $0.05$ (green), and $0.02$ (blue). }
\label{rtr_gam}
\end{figure}

Using equations (\ref{l_x_cor1})-(\ref{l_x_tot1}), we can calculate
the relation between the X-ray luminosity and spectral shape by
changing the value of transition radius $r_{\rm tr}$, if the values
of the mass accretion rate $\dot{m}$, the inner radius $r_{\rm in}$
of the ADAF, and $f_{\eta}^{\rm ADAF}$ are specified. In the
calculations, we assume that the photon spectral index of the hard
X-ray emission in 2$-$10~keV from the corona of the disc,
$\Gamma_{\rm cor}=2$, and $\Gamma_{\rm ADAF}=1$ for the X-ray
spectrum from the inner ADAF \citep*[][]{2012A&A...539A.104Y}. The
inner radius of the accretion disc $r_{\rm tr}=6$ is adopted in our
calculations. We plot the results with different values of the
parameters in Fig. \ref{eddrat_gam}, and the spectral indices
$\Gamma$ as functions of the transition radius $r_{\rm tr}$ are
plotted in Fig. \ref{rtr_gam}.

{It was suggested that the stationary transition radius may vary
with the mass accretion rate as
\begin{equation}
r_{\rm tr}=r_{\rm tr,0}\left({\frac
{\dot{m}}{\dot{m}_0}}\right)^{-p}, \label{r_tr_mdot1}
\end{equation}
where $p\sim 0.5-2$ for different accretion mode transition
mechanisms \citep*[see][for the detailed
discussion]{2004A&A...428...39C}. In Fig. \ref{eddrat_gam}, we plot
the results of the varying mass accretion rate model with $r_{\rm
tr,0}=r_{\rm in}$ and $\dot{m}_0=1$ adopted. It is found that the
result calculated with $p=2$ agrees quite well with the global
correlation between average spectral index and mass accretion rate
for the AGN sample. }

In our model, we assume that the transition radius between the ADAF
and the thin disc corona varies with time. We have to estimate the
timescale of variability caused by the variation of $R_{\rm tr}$.
The timescale of the transition radius moving toward the black hole
is comparable with the viscous timescale of the thin disc at $R_{\rm
tr}$, which can be estimated with
\begin{displaymath}
\tau_{\rm vis}\sim {\frac {R_{\rm tr}}{v_R}}={\frac {R_{\rm
tr}^2}{\alpha c_{\rm s}H}}
\end{displaymath}
\begin{equation}
=1.56\times 10^{-7}r_{\rm tr}^{3/2}\alpha^{-1}\left({\frac H
R}\right)^{-2}\left({\frac {M_{\rm bh}}{10^6{\rm
M}_\odot}}\right)~{\rm years}, \label{tau_vis1}
\end{equation}
where the viscosity parameter $\alpha<1$, and $H$ is the disc
thickness \citep*[][]{1973A&A....24..337S}. We note that the black
hole masses of four sources in \citet{2009MNRAS.399.1597S} are
around $10^6~M_\odot$, one is about $10^8~M_\odot$, and the
remainders are $\sim10^7~M_\odot$. We find that the black hole
masses of some sources in their work were overestimated. The updated
black hole mass measurements with broad line profiles for some
sources show that \citep*[see][for comparison]{2009MNRAS.399.1597S},
$1.06\times 10^8$~M$_\odot$ (Fairall), $7.13\times 10^5$~M$_\odot$
(NGC~4051), $3.19\times 10^6$~M$_\odot$ (NGC~3227),  $2.78\times
10^7$~M$_\odot$ (NGC~5548),  $1.24\times 10^7$~M$_\odot$ (NGC~3783),
and $1.32\times 10^7$~M$_\odot$ (NGC~3516)
\citep*[][]{2011MNRAS.412.2211G}; $7.0\times 10^6$~M$_\odot$
(NGC~5506) and $4.5\times 10^6$~M$_\odot$ (MCG-6-30-15)
\citep*[][]{2013MNRAS.430L..49M}. For typical values of the
parameters, for example, $r_{\rm tr}=100$, $\alpha=1$, {$H/R=0.05$
(corresponding to $\dot{m}=0.5$ at $r_{\rm tr}=100$) \citep*[e.g.,
see][]{2004MNRAS.355.1080L}}, and $M_{\rm bh}=10^7\rm M_\odot$, the
viscous timescale $\tau_{\rm vis}\sim 0.6$~year.

The cold disc is irradiated by the emission from the hot corona, and
the evaporation of the gas in the cold disc is assumed to be
balanced by the condensation of the gas in the corona in steady
cases. The disc corona will transit to an ADAF when the evaporation
dominates over the condensation. Thus, the typical timescale of the
transition from the disc corona to an ADAF is comparable with the
evaporation timescale. At the transition radius $R_{\rm tr}$, all
the gas in the cold disc is evaporated into the hot corona, and we
have
\begin{equation}
\pi R_{\rm tr}^2\dot{M}_{\rm evap}(R_{\rm tr})\sim \dot{M}=-2\pi
R_{\rm tr}\Sigma(R_{\rm tr}) v_{R}, \label{mdot_evap}
\end{equation}
where $\dot{M}_{\rm evap}$ is loss rate of the gas evaporated in
unit surface area of the disc, and $\Sigma$ is the surface density
of the disc \citep*[][]{1998PASJ...50L...5M}. The evaporation
timescale can be estimated as
\begin{equation}
\tau_{\rm evap}\sim {\frac {\Sigma}{\dot{M}_{\rm evap}}}=-{\frac
{R_{\rm tr}}{2v_{R}}}.  \label{tau_evap1}
\end{equation}
Comparing with equation (\ref{tau_vis1}), we find that the viscous
timescale of the disc is twice of the evaporation timescale.
Considering the estimates of the timescales being at order of
magnitude, a factor of two does not mean much. We simply adopt
$\tau_{\rm vis}$ as the typical timescale of the variability caused
by the variation of the transition radius $R_{\rm tr}$.

\section{Discussion}\label{discussion}
{After taking account of a constant reflection component in the
X-ray spectra\footnote
{the amplitude of the constant refection component was chosen to have $R=1$
relative to the time-averaged continuum.},
\cite{2009MNRAS.399.1597S} suggested the intrinsic X-ray photon}
index in individual AGNs varies as $\Gamma_{\rm intrinsic} \sim
F^{0.08}_{\rm 2-10 keV}$, consistent with the average spectral
index--accretion rate relation in their AGN sample. However, the
assumption that the reflection component, which is believed from the
accretion disc and the torus, does not respond to central continuum
variations over  $\sim$ 10 years is questionable. The measured
infrared lags in nearby Seyfert galaxies are in order of $\sim$ 10
days to one year \citep[][]{Suganuma2006}, indicating the reflection
component from the torus should respond to central X-ray continuum
variation within the observed timescale of
\cite{2009MNRAS.399.1597S}. The response of the disc reflection to
the central continuum is more complicated than a constant component
\citep{Iwasawa96,Wang99,Wang01,Shu10_NGC2992,Fabian02_MCG-6-30-15}.
Therefore, the intrinsic X-ray variation in individual AGNs should
follow a slope steeper than $\Gamma_{\rm intrinsic} \sim
F^{0.08}_{\rm 2-10 keV}$, and in this work,  assuming the reflection
component responds well to central X-ray emission, we compare the
binned spectral index-flux data with our model \citep*[see the data
points in Fig. 8 of ][]{2009MNRAS.399.1597S}.

The emission from the corona decreases rapidly with increasing
transition radius, and therefore the observed emission is
predominantly from the inner ADAF truncated at a large radius
$r_{\rm tr}$, of which the spectrum is harder than that of the
corona. Thus, we find that the spectral index decreases with flux
even if the mass accretion rate is fixed (see Fig.
\ref{eddrat_gam}). As a similar spectral index-flux correlation is
present for most of the sources in the sample
\citep*[][]{2009MNRAS.399.1597S}, we take the data of NGC~3516 as an
example for comparison with our model. We find that the model
calculation with a relative low $f_{\eta}^{\rm ADAF}=0.05$ can
roughly reproduce the observed spectral index-flux data (the green
line in Fig. \ref{eddrat_gam}). This is qualitatively consistent
with the prediction of the theoretical works that the radiation
efficiency of ADAFs is significantly lower than the standard thin
accretion discs \citep*[see][for detailed
discussion]{1998tbha.conf..148N}. The range of the observed
variability of $\Gamma$ can be reproduced when the transition radius
$r_{\rm tr}$ fluctuates between $r_{\rm in}$ and several hundreds,
if $f_{\eta}^{\rm ADAF}\ga 0.05$ (see Fig. \ref{rtr_gam}). This is
independent of the mass accretion rate $\dot{m}$.
It is found that the timescale of the X-ray variability caused by
the transition radius changing is at the order of one year for a
typical AGN as NGC~3516 (see the discussion in Sect. \ref{results}),
which can be as short as tens of days if $M_{\rm bh}\sim10^6{\rm
M}_\odot$. This is roughly consistent with the observed light curves
of these sources given in \citet{2009MNRAS.399.1597S}. {We find that
$L_{\rm X, 2-10keV}/L_{\rm Edd}=3.8\times 10^{-3}$ for NGC~3516,
which is the minimal one in the sample with updated black hole mass
data. This implies that the viscous timescales of other sources with
similar or smaller black holes should be shorter than that of
NGC~3516. For the sources with softer X-ray spectra, i.e.,
larger-$\Gamma$, most of the X-ray emission is from the corona, and
the transition radius $r_{\rm tr}$ should be smaller, which
corresponds a shorter viscous timescale. We find that $\dot{m}\ga
0.5$ is required to reproduce the observations, which is slightly
higher than that expected by the ADAF model even if $\alpha=1$
\citep*[e.g., $\dot{m}\la\dot{m}_{\rm crit}\sim 0.3\alpha^2$ for an
ADAF, see][for detailed discussion]{1998tbha.conf..148N}. The radial
velocity of a thin accretion disc with magnetic outflows can be much
higher than that of a normal viscously driven disc, if the angular
momentum of the disc is removed predominantly by the outflows
\citep*[][]{2013ApJ...765..149C}. This mechanism may work in the
ADAF with magnetic outflows, and therefore its density would be
lower than that predicted by the conventional ADAF model for a given
mass accretion rate, which may lead to a higher critical accretion
rate for accretion mode transition.} Our present calculations are
carried out with $r_{\rm in}=6$ for non-rotating black holes. The
radiation efficiency of an accretion disc surrounding a spinning
black holes is higher than the non-rotating black hole case. A
higher radiation efficiency for a spinning black hole can be
mimicked by tuning the value of $r_{\rm in}$. We find that the
results are similar to those reported in Figs. \ref{eddrat_gam} and
\ref{rtr_gam} even if a smaller $r_{\rm in}$ is adopted.

When the transition radius decreases, more gravitational power is
released in the corona, and therefore the X-ray emission from the
corona increases, while the emission from the inner ADAF decreases.
The total X-ray emission increases with decreasing transition
radius, because the radiation efficiency of the outer disc/corona is
higher than that of the inner ADAF. The fraction of the X-ray
emission from the corona to the ADAF increases with decreasing
radius, which leads to the observed X-ray spectral index increasing
with luminosity, as the X-ray emission from the corona is softer
than that of the ADAF. The emission from the thin disc, mostly in
optical/UV wavebands, is also expected to increases with X-ray
emission. This is consistent with observations that reveal generally
simultaneous long term optical/UV and X-ray variations in AGNs
\citep[e.g.][]{Gaskell2006}. {In this work, the X-ray spectral
indices of the ADAF and corona are taken as input model parameters,
whereas the spectral index of ADAF may decrease with increasing
transition radius due to the interception of the soft photons from
the outer disc by the ADAF \citep*[][]{1999MNRAS.303L..11Z}. If this
effect is included, less fluctuation in transition radius is
required to explain the observed X-ray variability. }

The timescale of the variability caused by the transition radius is
proportional to the black hole mass (see equation \ref{tau_vis1}).
This implies that only the sources with small black holes ($\la
10^7{\rm M}_\odot$) will exhibit violent variability in the
timescale of $\la 1$ year. For those sources containing massive
black holes ($\sim 10^{8-9}{\rm M}_\odot$), the timescale of such
variability can be as long as tens or hundreds of years, which may
be unpractical as monitoring objects.

{The black holes are fed by the circumnuclear gas, and the accretion
rate is predominantly regulated by the properties of the gas at the
outer radius of the disc. The timescale of accretion rate changes is
much longer than the X-ray variability timescale. So, the average
X-ray flux may correspond to the composite emission from the ADAF
and the corona connected at the equilibrium transition radius, which
may vary with the mass accretion rate as expected by different
physical mechanisms for accretion mode transition. The best-fitting
average spectral index-mass accretion rate correlation for the AGN
sample is roughly consistent with our calculations with $p=2$ (see
equation \ref{r_tr_mdot1}), which seems to support the thermal
instability triggering accretion mode transition
\citep*[][]{1995ApJ...438L..37A}. However, it should be cautious on
this issue, because the correlation is derived with a sample of only
ten sources. Further observation on a large sample of AGN may help
understand the physics of the accretion mode transition. }

\section*{Acknowledgments}
We thank the referee for his/her helpful comments/ suggestions. This
work is supported by the NSFC (grants 11173043, 11121062, and
11233006), the Strategic Priority Research Program ``the Emergence
of Cosmological Structures" of the CAS (grant No. XDB09000000), and
Shanghai Municipality.

\end{document}